\documentclass[twocolumn,english,aps,prl,groupedaddress,showpacs]{revtex4-1}
\usepackage[T1]{fontenc}
\usepackage[latin9]{inputenc}
\usepackage{textcomp}
\usepackage{graphicx}
\usepackage{times}
\usepackage{babel}

\begin{document}
\title{How super-tough gels break}
\author{Itamar Kolvin$^{1,2}$, John M. Kolinski$^{1,3}$, Jian Ping Gong$^{4}$ and Jay Fineberg$^{1}$}
\affiliation{$^{1}$The Racah Institute of Physics, The Hebrew University of Jerusalem, Jerusalem, 91904, Israel \\ $^{2}$UC Santa Barbara, Santa Barbara, California, 93106, USA \\ $^{3}$\'{E}cole Polytechnique F\'{e}d\'{e}rale de Lausanne, Lausanne, 1015, Switzerland \\ $^{4}$Faculty of Advanced Life Science and Soft Matter GI-CoRE, Hokkaido University, Sapporo, 001-0021, Japan}

\begin{abstract}
Fracture of highly stretched materials challenges our view of how things break. We directly visualize rupture of tough double-network (DN) gels at >50\% strain. During fracture, crack tip shapes obey a $x\sim y^{1.6}$ power-law, in contrast to the parabolic profile observed in low-strain cracks. A new length-scale $\ell$ emerges from the power-law; we show that $\ell$ scales directly with the stored elastic energy, and diverges when the crack velocity approaches the shear wave speed. Our results show that DN gels undergo brittle fracture, and provide a testing ground for large-strain fracture mechanics.
\end{abstract}

\pacs{}
\maketitle

Gels and rubbers are soft materials that are often tough to break \cite{creton2016,ducrot,filippidi}. Fracture in these materials typically occurs at strains exceeding 10\%. In contrast, our common understanding of how things break, Linear Elastic Fracture Mechanics (LEFM) \cite{griffith,irwin, eshelby, freund}, is a \textit{small} strain theory. 
LEFM predicts that under applied tension, stresses at a distance $r$ from a crack's tip diverge as $\sigma \sim 1/\sqrt{r}$, leading to a parabolic crack opening. Knowing the stress distribution allows one to calculate the energy flux into a crack's tip, $G$, which must balance the energetic cost of fracture $\Gamma$. Energy balance $G=\Gamma$ results in an equation of motion for cracks that predicts that crack velocities are limited by the Rayleigh wave speed $c_R$. Previous studies have shown that these predictions remain valid under moderate applied strains \cite{boue}. How fracture mechanics change for cracks propagating under large strains, where strongly nonlinear effects appear, is an unresolved question.

Tough hydrogels are a novel arena for testing theories for large strain fracture \cite{webber,nakajima,mayumi,sun}. Hydrogels are aqueous solids that owe their rigidity to a sparse network of cross-linked polymer chains \cite{tanaka1}. Their resemblance and compatibility with biological tissues make hydrogels natural candidates for biomedical applications \cite{lee,haque}. However, the low fracture toughness of common gels \cite{chen, zhao} $(\Gamma \sim 10 J/m^2)$ is a major limit to their performance.  By increasing the complexity of gel structure such limits may be pushed much further than previously thought \cite{tanaka2, wu}. In a double-network (DN) gel, following the polymerization of a first brittle network, the gel is immersed in a bath of a second monomer which is then polymerized to make a loosely cross-linked second network \cite{gong1}. The fracture energies $\Gamma$ of these materials are orders of magnitude higher than either of the individual networks on their own \cite{yu}. The origin of these remarkable properties lies in the ability of the material to dissipate a large portion of the mechanical work through dissociation of sacrificial bonds in the first (stiff) network, while the second (soft) network keeps the medium intact \cite{ahmed}.

Here we study the dynamic fracture of covalently cross-linked DN gels by directly visualizing how they fracture in real time. We observe the propagation of brittle cracks whose opening follows a $\sim\!\!\!\! 1.6$ power-law that deviates from the parabolic shape predicted by LEFM. In analogy to LEFM, we define a length-scale $\ell$ from the crack tip opening and show that it grows with the elastic energy stored in the sample prior to crack propagation. Hence, whereas dynamic cracks in DN gels do not strictly abide by LEFM, they bear striking similarities to cracks in brittle elastic materials. 

\begin{figure}[h]
	\protect\includegraphics[width=\linewidth]{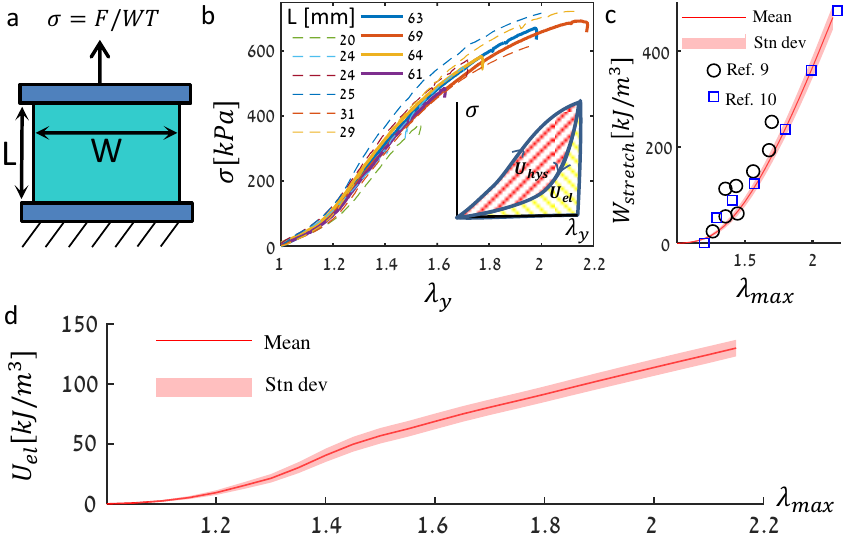}\caption{
		(a) A $T=1.3$mm thick rectangular sheet of DN gel is uniformly loaded by two rigid grips. A load-cell monitors the tension $F$ in the sheet.   (b) The nominal tensile stress versus the stretch along the $y$ axis for uniformly loaded samples prior to fracture. Colors: values of $L$ used. (inset) A schematic stress-stretch loop depicting the loading and unloading cycle. Shaded areas define the work dissipated in a cycle $U_{hys}$, and the stored elastic energy $U_{el}$ following the loading step. (c) The work density due to stretching $W_{stretch}$ as a function of the maximum stretch, $\lambda_{max}$.  $W_{stretch}$ was obtained by integrating the area under the average of the curves in (b). These measurements (red line) compare well with previous studies \cite{webber,nakajima}. Red shading denotes standard deviations. (d) The stored elastic energy $U_{el}=W_{stretch}-U_{hys}$ computed from (c) using published values of  $U_{hys}/W$  \cite{nakajima}. 
		\label{fig1}}
\end{figure}

Our experiments were conducted using DN hydrogels prepared according to the following protocol \cite{gong1,ahmed}. 
The first network was polymerized from a 1M solution of 2-acrylamido-2-methylpropane sulfonic acid sodium salt (NaAMPS) with 4mol\% N,N'-Methylenebisacrylamide (MBAA) as the cross-linker and 1mol\% $\alpha$-ketoglutaric ($\alpha$-keto) acid as the initiator. The second network was polymerized from a 2M solution of acrylamide with 0.01mol\% MBAA and 0.01mol\% $\alpha$-keto. 
The gels were prepared in a reaction cell consisting of two glass plates separated by a silicone spacer. 
Fully water-swollen gel sheets, of thickness $T=1.3$mm, were cut into a rectangular shape, of width $W=70-80$mm along the $x$ (propagation) dimension. 
We employed samples of both large ($L>60$mm) and small ($L<31$mm) aspect ratios in the loading ($y$) direction. Prior to each experiment, spray paint was applied to one face of the gel sheet to create a dense pattern for further processing by means of Digital Image Correlation (DIC) \cite{blaber}. Instantaneous displacements and strains applied to ``virgin'' (previously un-stretched) gel samples were thus determined during both the loading cycle and the fracture experiments in this manner.
The gel sheets were  placed into a custom, self-tightening gripping mechanism, and loaded in tension along the $y$-axis by a translation stage with a loading rate of 1 mm/sec, as depicted in Fig.~\ref{fig1}(a). The applied load $F$ was monitored at 100 Hz by an amplified load cell signal. To determine the instantaneous state of stretch, images of the gel sheet were taken at 3 frames per second during the loading.

Dynamic fracture occurs when the energy flux into the crack tip, $G$, is equal to the fracture energy, $\Gamma$ (the energy dissipated per unit crack length). Do DN gels behave as elastic media during dynamic fracture? To answer this question, we first determine the amount of elastic energy stored in the sample prior to fracture via the stress-stretch relation during the loading stage. Fig.~\ref{fig1}(b) presents the nominal stress $\sigma = F/TW$ as a function of the applied stretch $\lambda_y$, measured in-situ using DIC when each sample was stretched for the first time. The stress-stretch behavior of the material is reproducible and independent of sample size, as shown for 10 samples. Fig. \ref{fig1}(c) depicts the total work per unit volume $W_{stretch} (\lambda_{max})=\int_1^{\lambda_{max}}\left<\sigma(\lambda_y )\right>  d\lambda_y$, computed by averaging $\sigma(\lambda_y)$ over our data and integrating the areas under the average curve $\left<\sigma(\lambda_y )\right>$. The resulting values agree with previous studies \cite{webber,nakajima}. 
During the loading stage, a significant part of the stretching work is dissipated through dissociation of sacrificial bonds (see inset of Fig.~\ref{fig1}(b)), which we denote by $U_{hys}$. We estimate the remaining elastic energy per unit volume $U_{el}=W_{stretch}-U_{hys}$, presented in Fig.~\ref{fig1}(d), by using previously published values of the relative hysteretic loss $U_{hys}/W_{stretch}$ (50\%-70\% for the stretches considered here). Interestingly, in the few minutes between the cessation of loading and crack initiation, the stress in the sample relaxed  by a few percent, with no associated material displacement, resulting hook-shaped endings of the curves in Fig.~\ref{fig1} (b). This relaxation indicates that material damage is at least partially rate dependent \cite{yu}.

\begin{figure}[h]
	\protect\includegraphics[width=\linewidth]{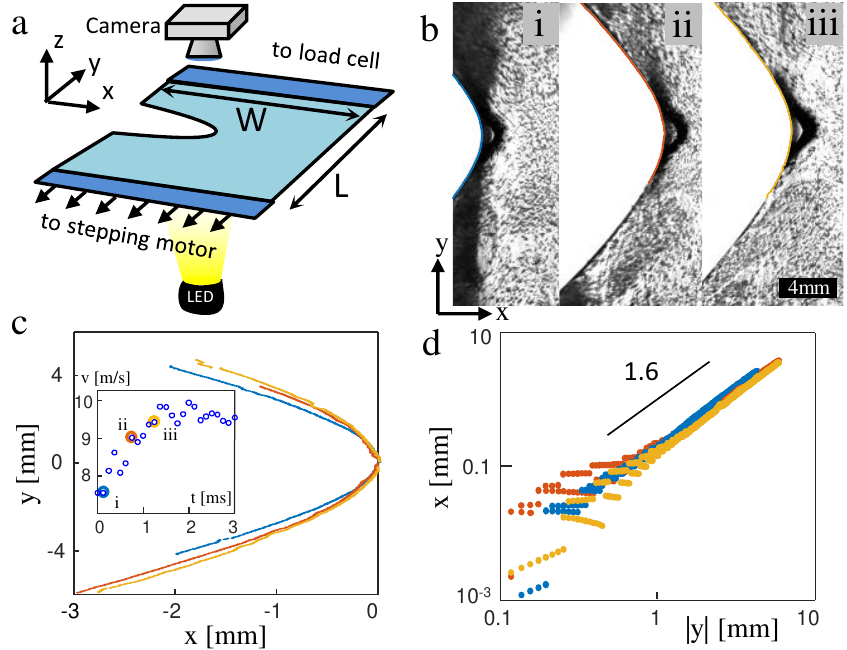}
	\caption{(a) Virgin gel samples are first uniformly stretched. A crack is then inserted at the middle of the sample's $y$-edge. Propagating cracks are imaged via a fast camera as shown. (b) Image sequence of an accelerating crack and (c) the CTODs extracted from each image and translated so that the crack tip is at the origin of axes. (inset) The velocity trace, where the colored symbols denote the frames i-iii in (b). (d) CTODs extracted from each image in (b) follow a $\sim\!\! 1.6$ power-law.}
	\label{fig2}
\end{figure}

Upon the cessation of loading, we used a razor blade to introduce a small notch into the edge of the pre-stressed sample along its centerline ($y=L/2$).
At a critical notch length, rapid crack propagation immediately ensued. 
Crack dynamics were recorded by a 2 megapixel high-speed camera at 8000 frames per second, in a window of $x \times y= 40 \times 20$ mm (in the laboratory frame) surrounding the center of the sample with a typical resolution of 20 $\mu$m per pixel. The transparent gels were illuminated (see Fig.\ref{fig2}(a)) by a 2 $\mu$sec-pulsed, spatially uniform and monochromatic LED light source.

\begin{figure}[b]
	\protect\includegraphics[width=\linewidth]{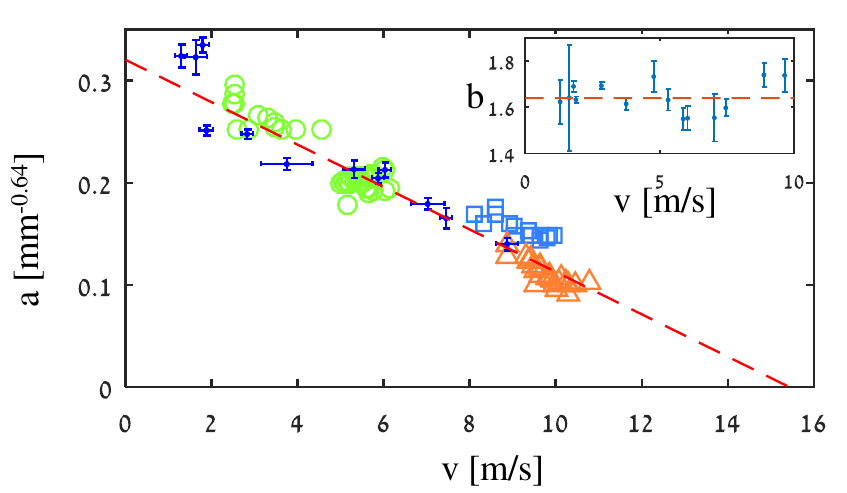}
	\caption{The prefactor $a$ extracted from power-law fits $x=a y^{1.64}$ over the range $|y|<3$mm to CTODs for: steadily propagating cracks (blue points are averages over >10 images at a given $v$, error-bars show standard deviations); accelerating cracks (green circles, blue squares, orange triangles --- each set of points represents a separate experiment; each point corresponds to a single image). A linear regression over all $a$ values intersects the $\mathrm{v}$ axis at $\mathrm{v}\simeq 15$m/s (red dashed line). (inset) Averages and standard deviations (error-bars) of the exponent $b$ when fitting CTODs of steady propagating cracks to $x=ay^b$. Each point corresponds to a single experiment. The dashed red line lies at the average $b=1.64\pm0.08$ over all experiments. }
	\label{fig22}
\end{figure}

The crack tip opening displacement (CTOD) provides a window into the material properties at the extreme strains for which fracture occurs. Our real-time imaging of propagating cracks allows us to directly study the CTOD \cite{goldman} and its dependence on loading conditions and crack velocities, $\mathrm{v}$. Extracting the CTOD from each image (Fig.~\ref{fig2}(b,c)) reveals that cracks in DN gels deviate significantly from the LEFM-predicted parabolic profile $x \sim y^2$ \cite{irwin}. Instead, the crack shape follows a $x \sim y^{1.6}$ power-law (Fig.~\ref{fig2}(d)). For a given velocity, we find no significant difference in the CTODs measured for accelerating cracks and cracks propagating at a steady velocity, suggesting that these cracks have no inertia \cite{goldman}. Fitting the power-law function $x=ay^b$ to CTODs at steady velocities, we find that $b$ does not depend on velocity, as shown in Fig.~\ref{fig2}(c,d) and Fig.~\ref{fig22}; averaging over >240 CTODs yields $b=1.64 \pm 0.08$, where the error is the standard deviation. We fix $b=1.64$ and extract the pre-factor, $a$, as a function of $\mathrm{v}$ (Fig. ~\ref{fig22}). $a$ \textit{decreases} approximately linearly with $\mathrm{v}$, indicating a systematically \textit{increasing} CTOD with $\mathrm{v}$. A linear regression intersects the velocity axis at $\mathrm{v} \simeq 15$m/s; this would correspond to a completely open crack (zero curvature) profile.

\begin{figure}[h]
\protect\includegraphics[width=\linewidth]{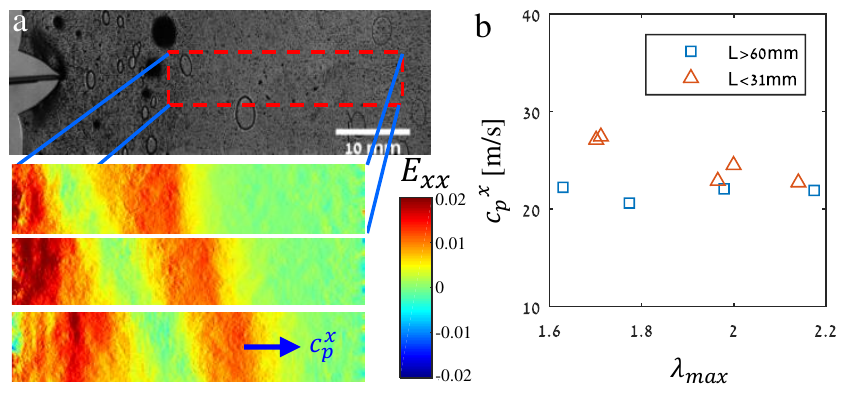}\caption{
In-situ determination of the longitudinal wave velocity along the $x$-axis, $c_p^x$. (a) At the initiation of fracture an elastic compressive wave propagates along the $x$ direction. DIC analysis of the region highlighted in red (top) is presented in three consecutive snapshots which yield a pulse-like $E_{xx}$ disturbance traveling at a constant velocity $c_p^x$ (bottom panels; the time interval between frames is 0.125 msec). (b) The variation of measured $c_p^x$ with the background stretch for different sample geometries (symbols). 
\label{fig3}}
\end{figure}

The nontrivial power law description of the CTOD defines a new dynamic length scale, $\ell=a^{-1/0.64}$. The measured range  $0.1 <a <0.35$ corresponds to $6 < \ell  < 32 $mm. In LEFM, the length scale defined by the CTOD, the radius of curvature, is proportional to $\Gamma/\mu$ where $\mu$ is the shear modulus. By analogy, we seek to determine whether $\ell$ is similarly related to the dissipation at the crack tip, or arises from another physical mechanism.  Relating $\Gamma$ to $\ell$ requires the determination of the high-rate shear modulus that is defined within the stretched (and damaged) state of the material. 

Fortunately, it is possible to directly measure the elastic response to a perturbation in-situ, immediately prior to fracture.  At the initial stages of crack propagation, a longitudinal elastic wave is emitted from the accelerating crack tip. This wave perturbs the strain field as it propagates along the $x$ direction, spanning the sample's breadth. Via DIC, we computed the instantaneous \textit{local} strain $E_{xx}$. Three sequential snapshots of typical strain fields formed by a wave propagating along the $x$-axis are presented in Fig.~\ref{fig3} (a). Averaging $E_{xx}(x,y,t)$ along the $y$-axis to reduce noise, we find that the pulse-like disturbance propagates at a constant speed $c_p^x$, corresponding to the plane stress longitudinal wave velocity along the $x$-axis in the lab frame. This measurement provides the value of $c_p^x$ at precisely the damaged and stretched conditions for which fracture takes place. Fig.~\ref{fig3} (b) depicts the variation of $c_p^x$ with the background stretch $\lambda_{max}$ and sample geometry $L$. $c_p^x \simeq 22$m/s is nearly constant for the large aspect ratio samples, while it is slightly larger and decreases with stretch in the small aspect ratio samples. 

\begin{figure}[h]
\protect\includegraphics[width=\linewidth]{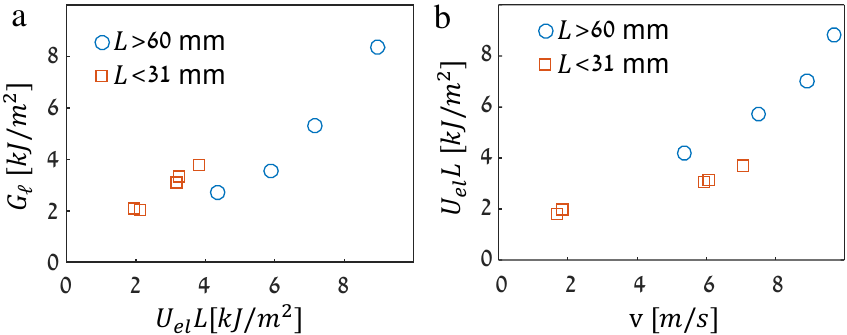}\caption{
Dissipation of energy in the dynamic fracture of DN gels. (a) The energetic measure, $G_\ell=\frac{1}{2}\rho c_p^2\ell$, approximates the stored elastic energy per unit area, $U_{el}L$, which is an upper bound for the energy dissipated in the fracture process. $U_{el}L$ is estimated from the quasi-static loading data in Fig. \ref{fig1}. (b) Growth of elastic energy dissipated in fracture with crack velocity.} 
\label{fig4}
\end{figure}

Is the length scale $\ell$ related to the dissipation inherent in fracture? We examine this question by studying the CTODs of cracks at steady velocities. When dissipation is confined to a microscopic zone surrounding the crack tip, cracks consume a fixed amount of elastic energy per unit area $G=\Gamma(v)$. In a small aspect ratio sample in uniform tension, G approximately equals the constant elastic energy stored per unit crack extension, $G\approx U_{el}L$ \cite{goldman}, and the crack is said to be under ``infinite strip'' conditions.  In a finite sample, the infinite strip approximation is valid $t_1 \sim L/c_s^y$ seconds after fracture initiation, where $c_s^y$ is the shear wave speed along the $y$-axis \cite{marder,petersan}.  By analogy with LEFM, we construct an energetic measure $G_\ell=2\mu \ell$ from the CTOD length-scale $\ell$. Here, the plane stress shear modulus is $\mu=\rho (c_p^x/2)^2$ \cite{shearmodulus,landau}, and $\rho \approx 1$g/cm$^3$ the gel density. $\mu \sim 120$kPa calculated from $c_p^x$ values in Fig.~\ref{fig3}(b) agrees with the small-strain modulus of the virgin samples (Supplementary Fig. 1) \cite{supp}.  

Plotting $G_\ell$ at steady velocities against $U_{el}L$ we observe (Fig.~\ref{fig4}(a)) that these energetic measures are nearly equal over a wide range of experimental conditions. A closer examination shows that while $G_\ell\simeq U_{el} L$ for small aspect ratio samples ($L<31$), $G_\ell\leq U_{el} L$  for aspect ratio 
$\sim 1 (L>60)$ samples. This difference might arise since the $L>60$ samples are ``strips'' only to a first approximation; in these samples the stored elastic energy is not fully dissipated in fracture. 

Brittle fracture occurs when inertia dominates dissipation and a crack, once formed, will continuously accelerate to acoustic speeds by releasing the elastic energy stored in the surrounding material. This contrasts with ductile fracture where crack propagation instantaneously mirrors external loads. Our results show that covalently cross-linked DN gels fail via brittle fracture. As Fig.~\ref{fig4}(a) shows, $\ell$ is proportional to the elastic energy released in fracture. Thus, $\ell$ plays a role analogous to the radius of curvature in parabolic LEFM CTODs. This correspondence indicates that the crack shape and the energy flux into the crack tip are mainly determined by elastic stresses in the material. Further support for this claim comes from Fig.~\ref{fig22} where the CTOD prefactor $a$ extrapolates to zero as $\mathrm{v}\rightarrow 15$m/s exactly like the crack tip curvature of LEFM cracks as $\mathrm{v} \rightarrow c_R$ \cite{freund,livne,bouchbinder}. 
For our incompressible gels, in plane stress, we estimate $c_R\!\sim\! c_p^x/2\!\sim\! 11$ m/s \cite{landau}, which is in the range of the extrapolated zero-crossing of $a$. Hence, we see that dynamic cracks in DN gels are similarly limited by acoustic speeds.

It is surprising that DN gels undergo brittle fracture while having very high $\Gamma$.
Typical values for hydrogels, $\Gamma \sim 10$J/m$^2$, are explained by the Lake-Thomas mechanism where polymer chains bridging the crack faces snap when stretched to their maximum length. 
Since for DN gels, $\Gamma \sim 100 - 1000$J/m$^2$ \cite{tanaka2} (Fig.~\ref{fig4}(b)) there must be an additional dissipative mechanism. 
In quasi-static tearing experiments \cite{yu}, the crack tip is surrounded by a `damage zone', whose $\sim$ mm size explains this substantial increase in fracture energy. 
The propagation of dynamic brittle cracks is only possible if dissipation is confined to a small region near the crack tip. 
Outside of this region, material response must be virtually elastic.
To gage whether a damage zone is present in our experiment, we measured the residual strains adjacent to a crack face in a fractured gel. 
A crack with $\ell=11$mm, and $U_{el} L\! \sim\! 4$ kJ/m$^2$ left behind a $h\! \sim\! 2$mm wide wake of $\sim\!\! 10$\% residual strains (see Supplementary Fig. 2)\cite{supp}.
Beyond $h$ residual strains are negligible. 
Wake formation may also be evident in the highly deformed bubble-like feature immediately ahead of the crack tips in Fig.~\ref{fig2}(b).
The clear separation of length scales $\ell \gg h$ therefore explains the brittle fracture observed in our experiments.

What happens if a crack is introduced into a rectangular DN gel sheet prior to stretching? 
At a constant loading rate, fracture initially progresses in a ductile manner (i.e. concomitantly with the loading) \cite{ahmed}. 
Preliminary results show, however, that at a certain length, the crack becomes unstable and rapid fracture ensues. 
Further study is necessary to elucidate the transition between the two regimes.

Given that material deformation surrounding dynamic cracks in DN gels is mostly elastic, why do CTODs have a 1.6 power-law? Non-parabolic CTODs of the form $x\!\sim\! |y|^b$ appear when the elastic energy density at large strains scales as $e(I)\sim I^n$, where $I$ is the first strain invariant \cite{geubelle,long}. A simple scaling argument shows that $b=2n/(2n-1)$ which results in $1<b<2$ for $n>1$ \cite{supp}. 
When $n>1$ the material stiffness $d\sigma/d\lambda$ increases with stretch. In rubbers, stiffening is observed due to the finite extensibility of polymer chains. To test the effect of rubber stiffening with strain on the CTOD, we solved for the deformation of a nonlinearly elastic Arruda-Boyce material \cite{boyce} surrounding a static crack using finite-elements and observed $x\!\sim\! |y|^b$ CTODs with $b<2$.
In DN gels, nonlinear elastic response is observed when they are loaded for a second time.  The stress-strain curve in a previously stretched sample, empirically, becomes increasingly steeper when approaching the previous maximal $\lambda_y$.  
Thus, nonlinear elastic stiffening at large strains may lead to the observed non-parabolic CTODs with $b<2$. 
Future work may probe how nonlinear elastic stiffening affects the CTOD exponent $b$ by modifying the gel network structure \cite{tanaka2,ahmed}.

While we expect that network stiffening with increasing strain will explain the power $b$, the CTOD should be also affected by the elastic anisotropy of the damaged gel. Our in-situ determination of $c_p^x$ reveals that $\mu$ is near the shear modulus of the undamaged material. It is therefore conceivable that while the first tensile loading changes the elasticity along the stretch axis drastically, it preserves the elasticity of the virgin material perpendicular to the stretch axis. This conclusion is consistent with observed anisotropic microscopic structure of damaged gels \cite{tominaga}. 

We have shown that brittle dynamic fracture ensues in pre-stretched DN gels at strains greater than 50\%. The crack tip profiles follow a nontrivial power-law that is characterized by a length that scales with the stored elastic energy divided by the shear modulus. We envision that our observations of the CTOD are closely linked to elastic properties of the gel, which are determined by complex and often hidden internal processes within the material. These macroscopic observations therefore provide a new window into the microscopic structure of the gel network under the extreme conditions that characterize fracture. We therefore believe that these results have the potential to both significantly improve our understanding of the origin of toughness in DN gels and pave the way for future theories of large strain fracture. 

Acknowledgments: J. F. and I. K. acknowledge the support of the Israel Science Foundation (Grant No. 1523/15), as well as the US-Israel Bi-national Science Foundation (Grant No. 2016950). J.M.K. acknowledges the Fulbright-Israel post-doctoral fellowship. J. P. G. acknowledges the support of ImPACT Program of Council for Science, Technology and Innovation (Cabinet Office, Government of Japan).

\bibliographystyle{apsrev}
\bibliography{dn}

\begin{thebibliography}{35}
\bibitem[{she()}]{shearmodulus}
\bibinfo{note}{The shear modulus is related to the shear wave speed via $\mu =
  \rho (c_s)^2$. In an incompressible elastic thin sheet, $c_s = c_p/2$
  \cite{landau}.}
\end{thebibliography}

\end{document}


\title{Supplementary material for ``How super-tough gels break''}
\author{Itamar Kolvin$^{1,2}$, John M. Kolinski$^{1,3}$, Jian Ping Gong$^{4}$ and Jay Fineberg$^{1}$}
\affiliation{$^{1}$The Racah Institute of Physics, The Hebrew University of Jerusalem, Jerusalem, 91904, Israel \\ $^{2}$UC Santa Barbara, Santa Barbara, California, 93106, USA \\ $^{3}$\'{E}cole Polytechnique F\'{e}d\'{e}rale de Lausanne, Lausanne, 1015, Switzerland \\ $^{4}$Faculty of Advanced Life Science and Soft Matter GI-CoRE, Hokkaido University, Sapporo, 001-0021, Japan}

\pacs{}
\maketitle

\section{Numerical values of the data presented in Fig. 5}
\begin{table}[htbp]
	\centering	
	\begin{tabular}{r|r|r|r|r}
		L [mm] & $\ell$ [mm] & $\lambda_y$ & $U_{el}L$ [kJ/m\textsuperscript{2}] & v [m/s] \\ \hline
		63.13 & 22.12 & 1.98  & 7.09  & 8.87 \\
		 69.01 & 35.14 & 2.17  & 8.9   & 9.64 \\
		 63.5  & 16.98 & 1.75 & 5.81  & 7.45 \\
		 61.44 & 11.35 & 1.69  & 4.28  & 5.31 \\
		 23.54 & 5.88  & 1.66 & 1.89  & 1.63 \\
	 23.36 & 5.54  & 1.69 & 2.08  & 1.8 \\
	 25.01 & 11.39 & 1.99  & 3.2   & 6.03 \\
	  30.53 & 12.12 & 1.94  & 3.12  & 5.87 \\
	   28.05 & 14.92 & 2.10 & 3.78  & 7.03 \\	 
	\end{tabular}
	\caption{The data used in Fig. 5}
\end{table}%

\pagebreak

\section{DN gel small strain response}

\begin{figure}[h]
	\includegraphics[scale=1]{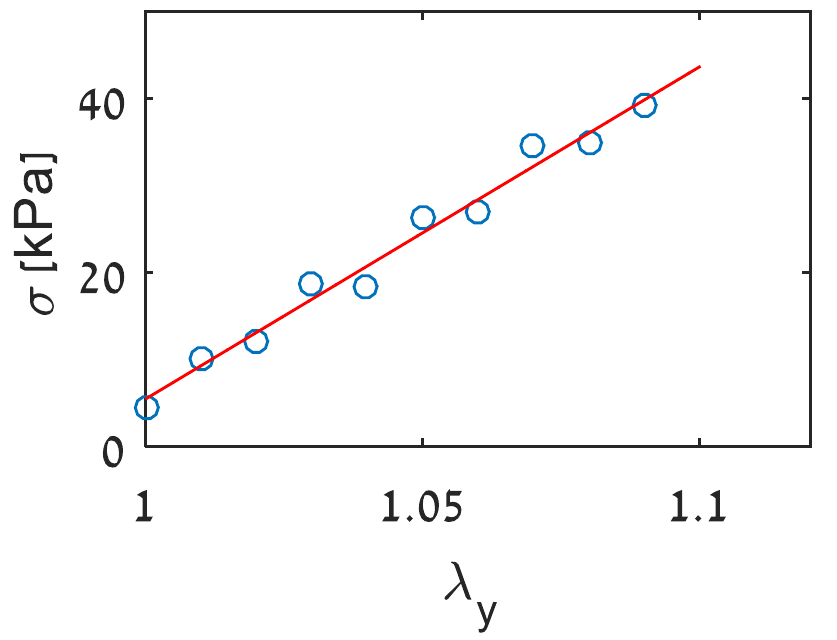}
	\caption{Stress-strain curve at small strains taken by averaging the curves of Fig. 1(b). The Young modulus $E=380\pm 40$ kPa is obtained via linear regression (red line). We estimate the shear modulus to be $\mu=E/3=130\pm 10$ kPa.}	
\end{figure}

\pagebreak

\section{Residual strains behind a propagating crack}
\begin{figure}[ht!]
	\protect\includegraphics[width=\linewidth]{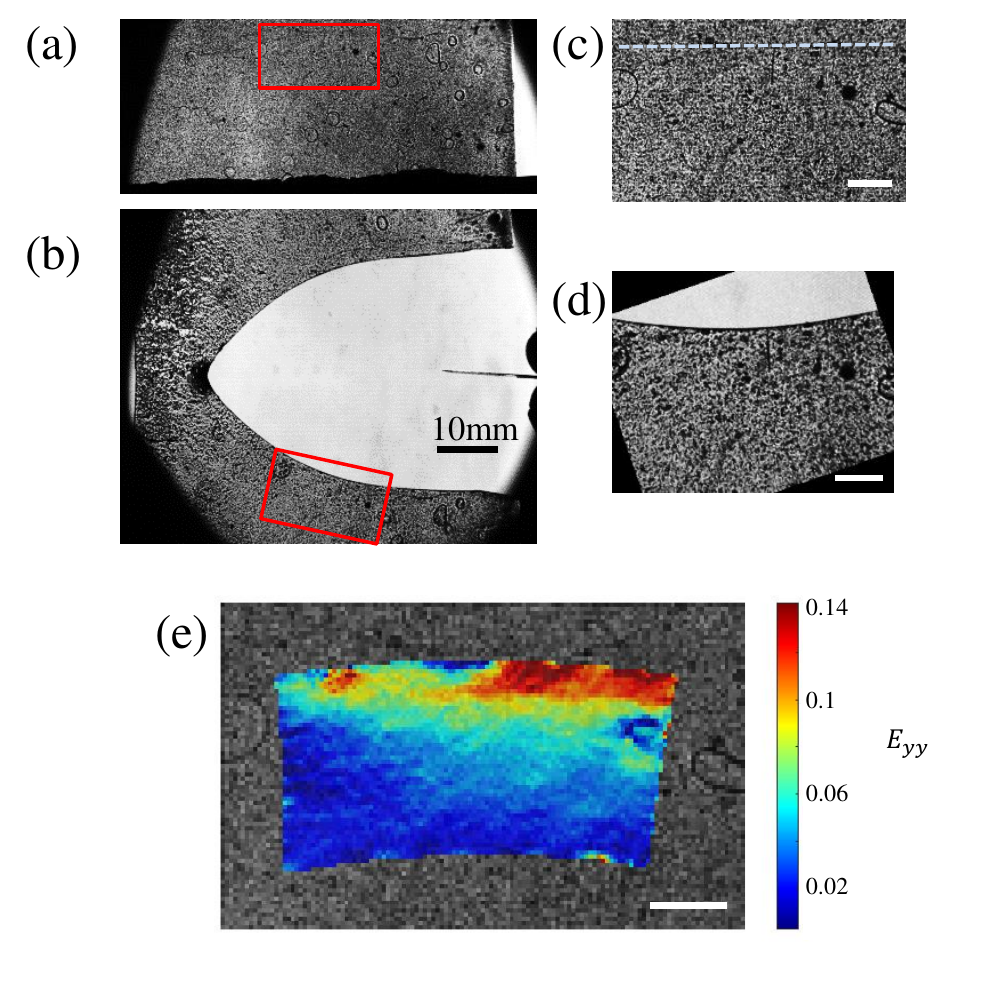}\caption{Residual strains within the tail of a running crack. (a) The gel sheet at its rest state. (b) A snapshot of the running crack under an imposed stretch of $\lambda_y=1.64$. (c,d) Zoomed sections highlighted in the red boxes in (a,b). The dashed line in (c) marks the approximate crack path in the non-stretched frame. (e) The residual $E_{yy}$ strain within the tail of the crack extracted through DIC of (c) and (d). White scale bars are 3 mm long.		
		\label{fig2}}
\end{figure}

\pagebreak[4]
%
\pagebreak
\section{CTOD scaling  at large strains}

In the main text we have observed that dynamic cracks in double-network gels have a CTOD of the shape $x\sim |y|^b$, where $b=1.64$. 
Here we give a heuristic argument for how non-parabolic CTOD's may arise due to large strain elasticity.

The elastic response of an isotropic material to deformation is derived from a strain energy density $e$, which depends on rotation invariants of the strain tensor. 
In 3D, there are three invariants \cite{marder2006}. Denoting the displacement of a point in the reference frame by $u_i$, and the strain tensor by $E_{ij} = \frac{1}{2}(\partial_j u_i +\partial_i u_j +\partial_i u_k\partial_j u_k)$, the first invariant is $I = E_{ii}$. Here $i$ and $j$ run over the three spatial coordinates $(X,Y,Z)$ in the material rest frame, and we use the Einstein summation convention.  We will assume that the contribution of the second invariant to $e$ is negligible, since it is second order in $E_{ij}$. The third invariant, which is related to volume changes, is a constant for incompressible materials (a good approximation for gels).
We therefore remain with $e= e(I)$. 
To obtain the CTOD scaling we note that one may construct a path independent integral around the crack tip, namely the J-integral \cite{rice,hutchinson}

\begin{equation}
J = \int_{\Gamma} \left(e(I)dy - S_{ij}n_j\partial_X u_i ds\right)
\end{equation} 
where $\Gamma$ is a contour surrounding a crack that lies along the negative $X$ axis, $S_{ij}$ is the nominal stress tensor, $n_j$ is the outward normal to the contour and $ds$ is the arc length element.
The path-independence of $J$ amounts to the existence of a $1/r$ singularity of the integrand as the radius $r=\sqrt{X^2+Y^2}$ approaches $0$. 
Since both terms in the integrand must scale in the same way ($S_{ij}\sim \partial e/\partial(\partial_j u_i)$), the energy density scales as $e(I)\sim 1/r$. 

For simplicity, let's assume the strain energy has a power-law dependence $e(I)\sim I^n$. 
For incompressible materials $\mathrm{det} (\delta_{ij}+\partial_i u_j) = 1$, which translates to $\partial_i u_i +\frac{1}{2}(\delta_{ij}\delta_{kl}\partial_i u_j \partial_k u_l - \delta_{ij}\delta_{kl}\partial_i u_k \partial_j u_l)+ \mathrm{det} \partial_i u_j=0$. The strain invariant is therefore $I = \partial_i u_i + (\partial_i u_k)^2 = (\partial_i u_k)^2- \frac{1}{2}(\delta_{ij}\delta_{kl}\partial_i u_j \partial_k u_l - \delta_{ij}\delta_{kl}\partial_i u_k \partial_j u_l)-\mathrm{det} \partial_i u_j$. 
The first term in $I$ is quadratic in the displacement gradient, and all others involve double and triple cross-products of displacement gradient components. 
Close to the crack faces, $\partial_X u_Y$ becomes the most dominant component, and thus $I\sim (\partial_X u_Y)^2$. 
Then,
\begin{equation}
(\partial_X u_Y)^{2n}\sim \frac{1}{r}
\end{equation}
 And hence $ u_Y \sim (-X)^{1-1/2n}$ or $ x\sim y^{2n/(2n-1)}$ where $x$ and $y$ are a lab frame coordinate system with origin at the crack tip, and $y=u_Y(X,Y=0,Z)$.
That is,
\begin{equation}
b = \frac{2n}{2n-1}\,.
\end{equation}

This expression for the CTOD power-law recovers the parabolic shape $b=2$ for the Neo-Hookean case $n=1$ \cite{boue} and approaches $b=1$ for large values of $n$. 
Our observed value of $b=1.64$ would arise in this model for $n=1.28$.
Long, Krishnan and Hui \cite{long} solved analytically the fracture problem in the context of the Generalized Neo-Hookean material model. This model is defined as $e(I) \sim \epsilon^{-1}\{(1+\epsilon n^{-1}(I-3))^n-1\}$, where $\epsilon$ and $n$ are parameters. This model approaches the Neo-Hookean behavior for small strains or small $\epsilon$. 
Their solution leads to a CTOD with $b=2n/(2n-1)$, the same as derived by the scaling argument above.

Both the power-law model $e(I)\sim I^n$ and the Generalized Neo-Hookean model are, to an extent, toy models for nonlinear elasticity. 
To test whether the CTOD scaling changes in more realistic models, we solved for the displacement fields surrounding a crack in an Arruda-Boyce material model using finite-elements software (Abaqus). 
Briefly, the Arruda-Boyce model is defined by $e(I)=\mu \lambda_m^2 f(\sqrt{I/3}/\lambda_m)$ where $\mu$ is an elastic modulus and the function $f\rightarrow \infty$ when $I\rightarrow 3\lambda_m^2$ (see \cite{boyce} for the definition of $f(\cdot)$). 
The latter property makes $\lambda_m$ the maximal stretch that the material can sustain.

We solved for the displacement fields surrounding a crack of length $35$mm in a rectangular sample of $L=65$mm and $W=70$mm.  We assumed that $\mu=100$kPa, and applied a constant displacement of $21.1$mm to the $y$ edge of the sample. Taking advantage of the symmetry of the problem, we solved it only in the upper half of the sample, assuming zero displacement along the $x$ axis, ahead of the crack tip. The solution used an adaptive triangular mesh of $\sim 800$ nodes.

\begin{figure}[!h]
	\protect\includegraphics[width=\linewidth]{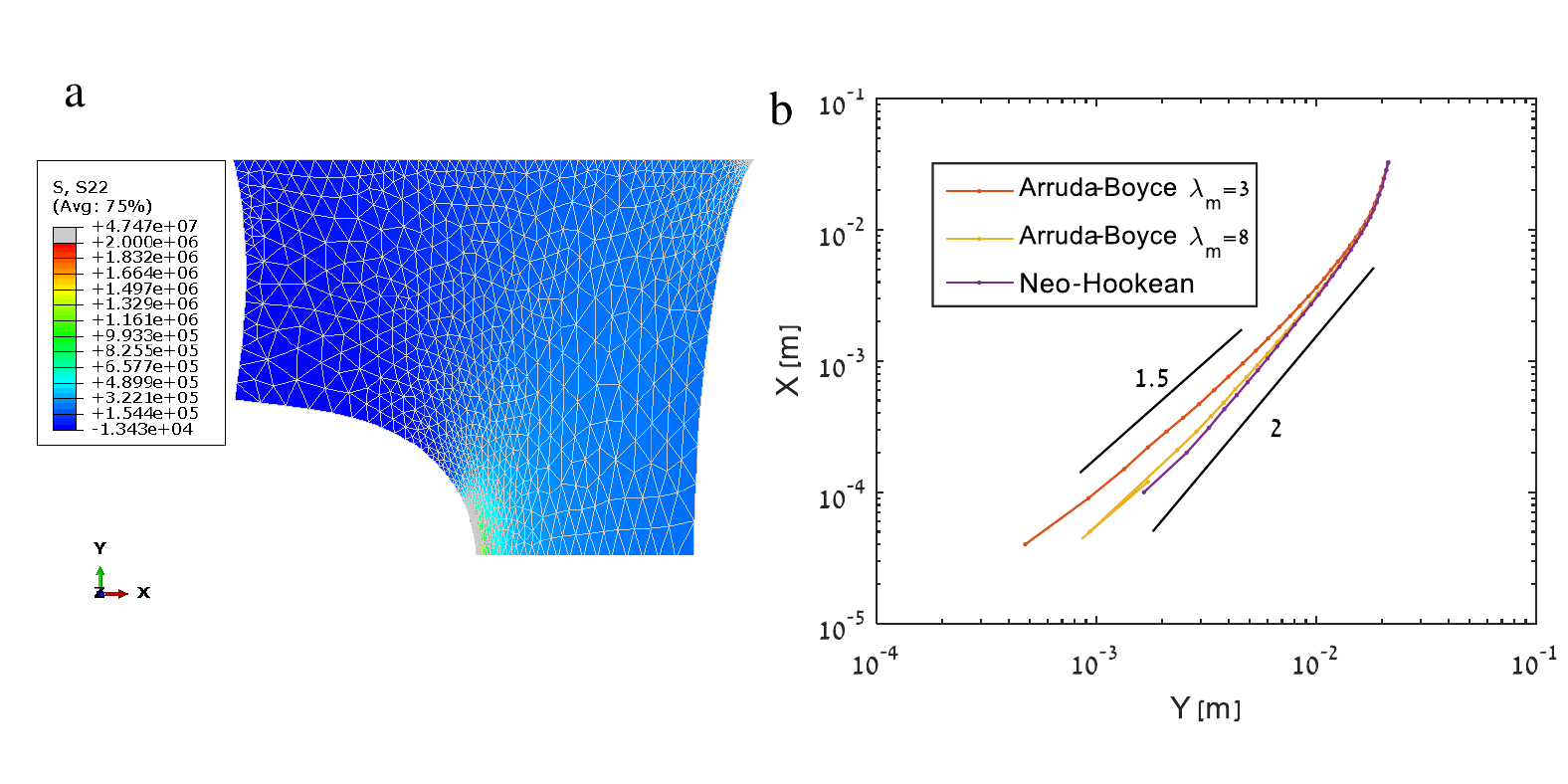}\caption{(a) The $S_{yy}$ stress component surrounding a crack in an Arruda-Boyce material. Only half of the sample is shown. (b) CTOD profiles for Neo-Hookean and Arruda-Boyce materials.
		\label{fig3}}
\end{figure}

Fig. \ref{fig3} shows an example of a solution and the CTOD profiles extracted from solutions obtained using three different materials. The Neo-Hookean model results in a parabolic CTOD, as predicted by the scaling argument. An Arruda-Boyce model with $\lambda_m=8$ shows a very small deviation from a parabola. Decreasing $\lambda_m$ to 3 results in a decrease of the CTOD power-law exponent to $\sim 1.5$.

In all of the models considered here, the exponent $b$ depends directly on the steepness of the stress-strain curve ($n$ in the power-law and Generalized Neo-Hookean models and $1/\lambda_m$ in the Arruda-Boyce model).
We anticipate that a future model that takes into account the anisotropy of our gels as well as nonlinear elasticity will be able to explain our observations.

\bibliographystyle{apsrev}
\bibliography{dn}